# Coherent ac spin current transmission across an antiferromagnetic CoO insulator

Q. Li[1+], M. Yang[1+], C. Klewe[2], P. Shafer[2], A. T. N'Diaye[2], D. Hou[3], T. Y. Wang[1], N. Gao[1], E. Saitoh[3], C. Hwang[4], R. J. Hicken[5], J. Li[6], E. Arenholz[2], and Z. Q. Qiu[1*]

[1]Department of Physics, University of California at Berkeley, Berkeley, California 94720, USA
[2]Advanced Light Source, Lawrence Berkeley National Laboratory, Berkeley, California 94720, USA
[3]WPI Advanced Institute for Materials Research, Tohoku University, Sendai 980-8577, Japan
[4]Korea Research Institute of Standards and Science, Yuseong, Daejeon 305-340, Korea
[5]Department of Physics and Astronomy, University of Exeter, Stocker Road, Exeter, Devon EX4 4QL, United Kingdom
[6]International Center for Quantum Materials, School of Physics, Peking University, Beijing 100871, China

**Abstract**

The recent discovery of spin-current transmission through antiferromagnetic (AFM) insulating materials opens up unprecedented opportunities for fundamental physics and spintronics applications. The great mystery currently surrounding this topic is: how could THz AFM magnons mediate a GHz spin current? This mis-match of frequencies becomes particularly critical for the case of coherent ac spin-current, raising the fundamental question of whether a GHz ac spin-current can ever keep its coherence inside an AFM insulator and so drive the spin precession of another FM layer coherently? Utilizing element- and time-resolved x-ray pump-probe measurements on Py/Ag/CoO/Ag/Fe$_{75}$Co$_{25}$/MgO(001) heterostructures, we demonstrate that a coherent GHz ac spin current pumped by the permalloy (Py) ferromagnetic resonance (FMR) can transmit coherently across an antiferromagnetic CoO insulating layer to drive a coherent spin precession of the FM Fe$_{75}$Co$_{25}$ layer. Further measurement results favor thermal magnons rather than evanescent spin waves as the mediator of the coherent ac spin current in CoO.

**Introduction**

Antiferromagnetic (AFM) materials have emerged as promising candidates for spintronic technology [1,2,3]. In particular, the discovery of spin-current transmission through AFM insulators [4,5,6,7,8,9] promotes their potential use for local spin switching within magnetic devices [10,11]. It is believed that the spin current propagation in AFM insulators is governed by THz magnons [12], which poses a great challenge for coherent GHz ac spin-current injection e.g., by ferromagnetic resonance (FMR) [13]. Due to the absence of GHz magnons in most AFMs, coherent GHz spin currents in AFM insulators have only been discussed in terms of evanescent waves [14] with other theoretical models [15,16,17] averaging out the THz ac components to focus on the dc spin current. This model is adequate to describe incoherent spin-current injection (e.g., spin Seebeck effect [6,7]), where only dc spin currents are observed. However, it raises fundamental questions for coherent ac spin-current injection and transmission (e.g., by FMR), where the frequency range (~GHz) is significantly lower than typical AFM magnon frequencies (~THz). Although FMR damping measurements indicate the injection of a GHz coherent ac spin current into a AFM layer [4,18,19,20,21], direct pump-probe measurements reveal that coherent magnons in AFM insulators can only carry net spins in the THz range [22]. Therefore, it becomes critically important to answer the question whether or not a GHz coherent spin current can propagate coherently across an AFM insulator to drive a coherent spin precession of another FM layer. Here, we report on experimental investigations of spin pumping, propagation, and transmission of a coherent GHz spin current in Py/Ag/CoO/Ag/$Fe_{75}Co_{25}$/MgO(001) using element- and time-resolved X-ray Magnetic Circular Dichroism (XMCD) and X-ray Magnetic Linear Dichroism (XMLD) [23,24].

**Results**

**Sample preparation and characterization.** Two samples of Py/Ag$^{(1)}$/CoO/Ag$^{(2)}$/$Fe_{75}Co_{25}$ were grown on top of MgO(001) substrates using Molecular Beam Epitaxy [Supplementary Fig. 1] with layer thicknesses of 30nm Py ($Ni_{20}Fe_{80}$), 2.5nm CoO, 5nm $Fe_{75}Co_{25}$, and 2nm Ag between Py and CoO for both samples. The 2nm Ag$^{(1)}$ between Py and CoO in these two samples permits a non-zero Py/CoO magnetic interlayer coupling which is important for the CoO spin alignment and for the Py spin pumping into the CoO [Ref. 14-17]. The Ag$^{(2)}$ thickness ($d_{Ag}$) between the CoO and the $Fe_{75}Co_{25}$ layers was varied from 2nm to 10 nm in the two samples to control the CoO/$Fe_{75}Co_{25}$ magnetic interlayer coupling [25], leading to the presence and absence of an equivalent interlayer coupling between the Py and $Fe_{75}Co_{25}$ magnetizations across the Ag/CoO/Ag spacer [Supplementary Fig. 3]. For convenience, we will refer to these two samples as Py/Ag/CoO/Ag(2nm)/$Fe_{75}Co_{25}$ and Py/Ag/CoO/Ag(10nm)/$Fe_{75}Co_{25}$, respectively.

DC XMLD measurements reveal a perpendicular coupling between the Py spins and the CoO AFM spin axis [26] (Fig. 1**a**) with a CoO Néel temperature of ~280 K [Supplementary Fig. 2]. We first performed FMR measurements using conventional power absorption detection to characterize the FMR resonance fields. The results show distinct Py and $Fe_{75}Co_{25}$ resonance fields [Fig. 1**b**] with the $Fe_{75}Co_{25}$ FMR disappearing below 8 GHz. The distinctly different Py and $Fe_{75}Co_{25}$ resonance fields enable us to selectively excite the Py pump layer at 4GHz and separately detect the spin current induced excitation of the $Fe_{75}Co_{25}$ sink layer (dashed line in Fig. 1**b**) using x-ray detected FMR [see Methods Section]. In addition, hysteresis loop and FMR results confirm the presence and absence of Py/Fe$_{75}$Co$_{25}$ interlayer coupling across the Ag/CoO/Ag(2nm) and Ag/CoO/Ag(10nm) spacer layers, respectively [Supplementary Fig. 3]. This allows us to separate the effect of the spin current from that of the interlayer coupling in driving the Fe$_{75}$Co$_{25}$ spin precession.

**Ac spin current transmission through the CoO layer.** Using element- and time-resolved XMCD measurements, we measured the Py and Fe$_{75}$Co$_{25}$ spin precession at the Py FMR field for a 4 GHz rf excitation. At T=280 K, the Co ac XMCD signal [Fig. 1**c**] clearly shows coherent Fe$_{75}$Co$_{25}$ spin precession in both Py/Ag/CoO/Ag(2nm)/Fe$_{75}$Co$_{25}$ and Py/Ag/CoO/Ag(10nm)/Fe$_{75}$Co$_{25}$ samples. Due to the presence and absence of the interlayer coupling, the observed Fe$_{75}$Co$_{25}$ spin precession can be attributed to different mechanisms for the two samples. In Py/Ag/CoO/Ag(2nm)/Fe$_{75}$Co$_{25}$ both the Py/Fe$_{75}$Co$_{25}$ interlayer coupling and the ac spin current contribute to the Fe$_{75}$Co$_{25}$ spin excitation, while it is dominated by the pure ac spin current in the Py/Ag/CoO/Ag(10nm)/Fe$_{75}$Co$_{25}$ sample. This assertion is supported by the observation of different amplitudes and phase delays of the Fe$_{75}$Co$_{25}$ spin precession in the two samples. At 180 K, *i.e.*, 100K below the T$_N$ of the CoO layer, we observe ~10% and ~60% decrease of the Fe$_{75}$Co$_{25}$ spin precession amplitude in Py/Ag/CoO/Ag(2nm)/Fe$_{75}$Co$_{25}$ and Py/Ag/CoO/Ag(10nm)/Fe$_{75}$Co$_{25}$ [Fig. 1**c**], respectively, confirming the presence of two different mechanisms driving the Fe$_{75}$Co$_{25}$ spin precession. To separate the interlayer coupling and the spin-current contributions, we measured the temperature-dependence of the Fe$_{75}$Co$_{25}$ precession amplitude normalized to the Py precession amplitude ($A_{FeCo}/A_{Py}$). In other words, the response of the spin sink layer is normalized to the strength of the spin source. The results [Fig. 1**e**] show that the $A_{FeCo}/A_{Py}$ values in both samples exhibit a broad peak around the CoO Néel temperature of 280K, similar to the behavior observed for dc spin-currents [6,7]. The difference in $A_{FeCo}/A_{Py}$ between these two samples is a temperature-independent constant, indicating that this difference is due to the Py/Fe$_{75}$Co$_{25}$ interlayer coupling in Py/Ag/CoO/Ag(2nm)/Fe$_{75}$Co$_{25}$ and that the common broad peak near 280K is due to transmission of a coherent ac spin current through the CoO layer. We then measured the Co spin precession in a Py(30nm)/Ag(2nm)/Co(1nm)/CoO(2.5nm)/MgO(001) reference sample [Fig. 1**f**] in which the Co(1nm) serves as an indicator of the spin precession driven by the Py FMR before the ac spin-current propagates through the CoO. The result shows a monotonic temperature dependence of the Co spin precession amplitude, suggesting that the broad peak in the Fe$_{75}$Co$_{25}$ spin precession amplitude near the CoO(2.5nm) Néel temperature of 280K is caused by the transmission of the ac spin current through the CoO. The temperature dependence of the Co spin precession amplitude in Fig. 1**f** is an interesting topic for future research, but is not the focus of the present work.

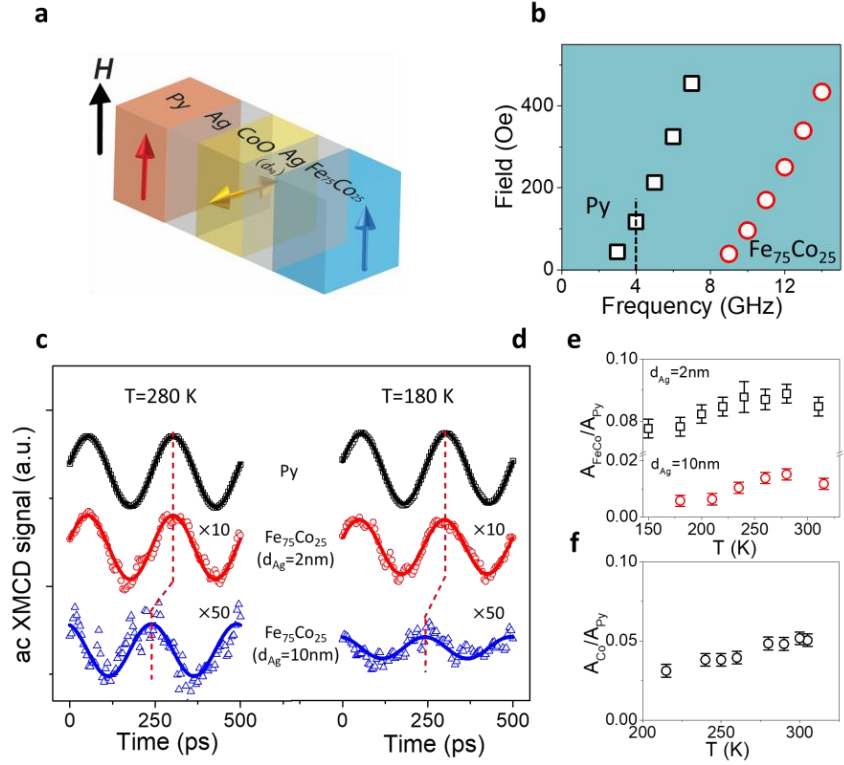

**Fig. 1| Ac spin current transmission through the CoO layer. a**, Schematic drawing of the spin configuration within the sample. **b**, FMR fields for the Py and $Fe_{75}Co_{25}$ layers within the Py/Ag/CoO/Ag(10nm)/$Fe_{75}Co_{25}$/MgO(001) sample. The dashed line shows where the ac XMCD/XMLD measurement was performed at the ALS. **c,d**, Ac XMCD signals showing Py and $Fe_{75}Co_{25}$ spin precession at the Py FMR field at 280 K and 180 K, respectively. **e** Temperature-dependent ratio of $Fe_{75}Co_{25}$ precession amplitude to Py precession amplitude $A_{FeCo}/A_{Py}$ for the Py/Ag/CoO/Ag(2nm)/$Fe_{75}Co_{25}$/MgO(001) and Py/Ag/CoO/Ag(10nm)/$Fe_{75}Co_{25}$/MgO(001) samples. **f** Temperature-dependent ratio of the Co precession amplitude to the Py precession amplitude for the Py/Ag/Co/CoO/MgO(001) sample.

**Separation of spin-current and interlayer-coupling.** Next we measured the Py and $Fe_{75}Co_{25}$ spin precession for different magnetic fields above and below the Py FMR. Both the amplitude and phase of the Py and $Fe_{75}Co_{25}$ spin precession were extracted by fitting the ac XMCD phase delay scans with a sinusoidal function. The Py spin precession amplitude exhibits the Lorentzian shape $A_{Py}^2 \sim \Delta H^2 / [(H - H_{res})^2 + \Delta H^2]$ expected for FMR, while the phase varies as $tan\varphi_{Py} = \Delta H / (H - H_{res})$ [red lines in Fig. 2**b,c,d,e**], exhibiting a total phase shift of 180° as the field is swept through the resonance, where $H_{res}$, $\Delta H$, $A_{Py}$, and $\varphi_{Py}$ are the Py FMR field, FMR linewidth, spin precession amplitude, and phase of precession, respectively. The $Fe_{75}Co_{25}$ spin precession amplitude $A_{FeCo}$ [Fig. 2**b,d**] exhibits a clear peak at the Py FMR field in both the Py/Ag/CoO/Ag(2nm)/$Fe_{75}Co_{25}$ and Py/Ag/CoO/Ag(10nm)/$Fe_{75}Co_{25}$ samples. Since $Fe_{75}Co_{25}$ does not undergo FMR at 4GHz, the peak in the $Fe_{75}Co_{25}$ precession amplitude at the Py FMR field proves that the enhanced Fe$_{75}$Co$_{25}$ spin precession must be induced by precession of the Py spins. To identify the different driving mechanisms in the two samples [Fig. 2**a**], we analyzed the Fe$_{75}$Co$_{25}$ precession phase $\varphi_{FeCo}$. The Fe$_{75}$Co$_{25}$ precession phase in the Py/Ag/CoO/Ag(2nm)/Fe$_{75}$Co$_{25}$ sample exhibits a monotonic field dependence, similar to that of the Py layer [Fig. 2**c**]. This can be understood as being the result of the Py/Fe$_{75}$Co$_{25}$ interlayer coupling, which favors parallel alignment of the Fe$_{75}$Co$_{25}$ and Py spins (both dc and ac components). In contrast, the phase of the Fe$_{75}$Co$_{25}$ precession in the Py/Ag/CoO/Ag(10nm)/Fe$_{75}$Co$_{25}$ sample [Fig. 2**e**] exhibits a clear bipolar behavior which is a fingerprint of spin-current driven precession [21,23].

To obtain a detailed quantitative understanding of the different mechanisms contributing to the Fe$_{75}$Co$_{25}$ precession, we consider the Landau-Lifshifts-Gilbert equation [21,23], where the quantities driving the precession include the rf field ($\vec{h}_{rf}$) from the coplanar waveguide (CPW), the effective Py/Fe$_{75}$Co$_{25}$ interlayer magnetic coupling energy ($-J_{int}\vec{m}_{Py} \cdot \vec{m}_{FeCo}$), and an ac spin current generated by the Py FMR ($\alpha_{Py}^{sp}\vec{m}_{Py} \times \frac{d\vec{m}_{Py}}{dt}$), where $\alpha_{Py}^{sp}$ is the Py spin pumping coefficient, and $\vec{m}_{Py}$ and $\vec{m}_{FeCo}$ are the Py and Fe$_{75}$Co$_{25}$ magnetization unit vectors, respectively. The important difference between the interlayer coupling and the spin current mechanisms is that their associated driving torques have a 90° phase difference, resulting in a distinctly different phase behavior for the Fe$_{75}$Co$_{25}$ spin precession (see below). As compared to the Fe$_{75}$Co$_{25}$ spin precession amplitude ($A_{FeCo}^0$) and phase ($\varphi_{FeCo}^0$) driven by $\vec{h}_{rf}$ only, the modification of the Fe$_{75}$Co$_{25}$ spin precession amplitude and phase by the interlayer coupling and spin current can be described by [Supplementary part 3]

$$\left|\frac{A_{FeCo}}{A_{FeCo}^0}\right| = \sqrt{1 + (\beta_{int}^2 + \beta_{sc}^2)sin^2\varphi_{Py} + 2\beta_{int}sin\varphi_{Py}cos\varphi_{Py} + 2\beta_{sc}sin^2\varphi_{Py}} \quad (1)$$

$$\tan(\varphi_{FeCo} - \varphi_{FeCo}^0) = \frac{\beta_{int}sin^2\varphi_{Py} - \beta_{sc}sin\varphi_{Py}cos\varphi_{Py}}{1 + \beta_{int}sin\varphi_{Py}cos\varphi_{Py} + \beta_{sc}sin^2\varphi_{Py}} \quad (2)$$

where $\beta_{int} = \frac{M_{Py}t_{Py}}{M_{FeCo}t_{FeCo}} \cdot \frac{\gamma J_{int}}{\alpha_{Py}\omega}$ and $\beta_{sc} = \frac{M_{Py}t_{Py}}{M_{FeCo}t_{FeCo}} \cdot \frac{\alpha_{Py}^{sp}}{\alpha_{Py}}$ correspond to the interlayer-coupling and spin-current mechanisms, respectively. Eqn. (1) shows that both $\beta_{int}$ and $\beta_{sc}$ enhance the Fe$_{75}$Co$_{25}$ precession amplitude by generating a peak in $A_{FeCo}$ in the vicinity of the Py resonance field, making it difficult to distinguish the spin-current and the interlayer-coupling effects purely by considering the precession amplitude. In contrast, eqn. (2) shows that only the spin current term ($\beta_{sc}$) leads to a bipolar phase behavior in the vicinity of the Py FMR ($\varphi_{Py} \sim 90°$) with $\varphi_{FeCo} > \varphi_{FeCo}^0$ for H < $H_{Py,res0}$ and $\varphi_{FeCo} < \varphi_{FeCo}^0$ for H > $H_{Py,res0}$. With $\varphi_{Py}$ derived from the Py data (red lines in Fig. 2**c,e**), we fit the Fe$_{75}$Co$_{25}$ spin precession amplitude and phase using eqn. (1) and (2) with $\beta_{int}$ and $\beta_{sc}$ as fitting parameters. The results agree very well with the experimental data (blue lines in Fig. 2**b-e**). Fig. 2**f** summarizes the $\beta_{int}$ and $\beta_{sc}$ values obtained for different temperatures. The interlayer coupling term $\beta_{int}$ has a finite value in Py/Ag/CoO/Ag(2nm)/Fe$_{75}$Co$_{25}$ but is virtually zero in Py/Ag/CoO/Ag(10nm)/Fe$_{75}$Co$_{25}$. The spin-current term $\beta_{sc}$ exhibits similar values in both samples with a broad peak around the CoO Néel temperature. This result clearly shows the different contributions made by the interlayer coupling and the spin current to the Fe$_{75}$Co$_{25}$ spin precession in the two samples. Based on these observations, the enhancement of the Fe$_{75}$Co$_{25}$ spin precession around T=280 K [Fig. 1**e**] can be attributed to an increase in the transmission of coherent ac spin current through the CoO layer around the Néel temperature. It is noteworthy that the equivalent Py/Fe$_{75}$Co$_{25}$ interlayer coupling across the Ag/CoO/Ag spacer shows relatively weak temperature dependence, even above the CoO Néel temperature. This may be an interesting basis for future studies [27], but it is not the focus of the present work.

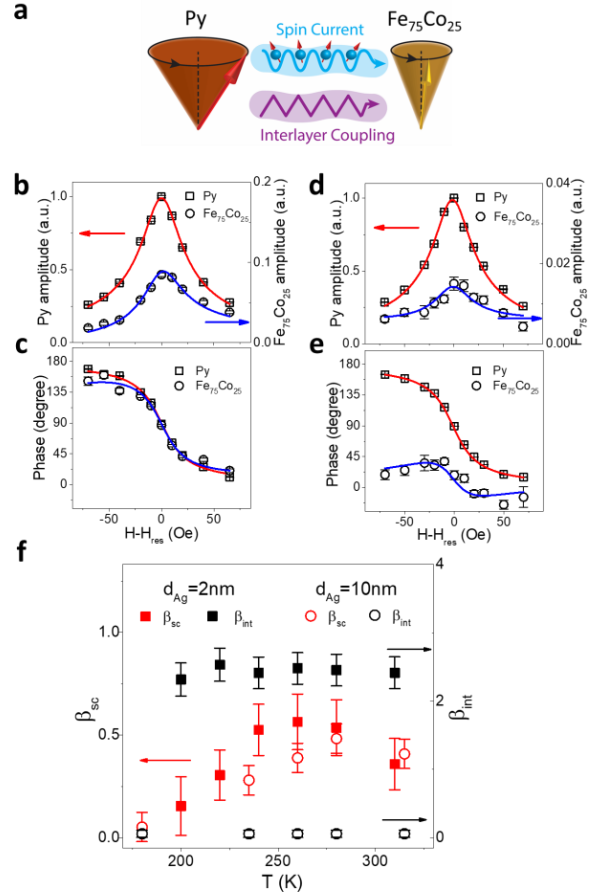

**Fig. 2| Separation of spin-current and interlayer-coupling contributions to the torque driving the Fe$_{75}$Co$_{25}$ spin precession. a**, Schematic drawing of ac spin current and interlayer coupling as the driving mechanisms of Fe$_{75}$Co$_{25}$ spin precession originating from the Py FMR. **b,d** Field-dependence of the amplitude and **c,e** field-dependence of the the phase of Py and Fe$_{75}$Co$_{25}$ spin precession at 280 K from **b,c** Py/Ag/CoO/Ag(2nm)/Fe$_{75}$Co$_{25}$ and **d,e** Py/Ag/CoO/Ag(10nm)/Fe$_{75}$Co$_{25}$, respectively. Red and blue lines are fits to the Py and Fe$_{75}$Co$_{25}$ signals, respectively. **f** Temperature-dependence of the spin-current ($\beta_{sc}$) and interlayer-coupling ($\beta_{int}$) coefficients for the two samples.

**Probing the AFM CoO spin precession**. To answer the question of whether the coherent ac spin current is carried by evanescent waves within the AFM [14], which involves a coherent GHz precession of the AFM spin axis, we prepared a sample of Py(30nm)/Ag(2nm)/CoO(2.5nm)/MgO(001) and performed ac XMLD measurements of the dynamics of the CoO moments. In detail, linearly polarized x-rays at normal incidence, with polarization axis tilted by 45° with respect to the CoO AFM spin axis, were utilized to detect the dynamic XMLD from the CoO layer. A rf evanescent GHz spin waves are excited within the CoO. However, our results show no detectable CoO ac XMLD signal both at 210 K and 280 K even with a data accumulation time much greater than that for the ac XMCD measurement (Fig. 3). Taking into account the noise level of the CoO ac XMLD signal, we estimate an approximate upper limit of the ac to dc signal ratio (ac XMLD divided by dc XMLD) to be less than 0.00005 for the CoO precession at the Py FMR field. For comparison, the ac XMCD/dc XMCD ratio of Py and $Fe_{75}Co_{25}$ in Py/Ag/CoO/Ag(10nm)/$Fe_{75}Co_{25}$ at the Py FMR field are estimated to be 0.011 and 0.00065, respectively. Therefore, the absence of the CoO ac XMLD signal suggests the absence of CoO evanescent modes in our sample, but rather promotes the idea of spin current transmission mediated by thermal magnons.

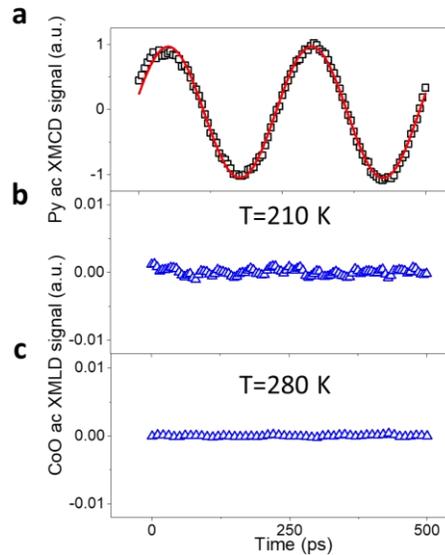

**Fig. 3| Measurement of AFM CoO spin axis precession. a**, Py ac XMCD signal at 210 K. CoO ac XMLD signal at **b** 210 K and **c** 280 K.

**Discussion**

It remains to be determined how THz thermal magnons can transport a coherent GHz spin current. It has been shown that a dc spin current is not carried by a single AFM thermal magnon mode. Instead, by lifting the population degeneracy of the left- and right-handed magnon modes of the AFM, i.e., by creating a right-handed magnon while annihilating a left-handed magnon of the same energy, a net spin angular momentum can be induced along the AFM spin axis without altering the energy state of the system considerably [12]. If GHz precession is considered to be an adiabatic process for the THz AFM magnons, then it may be possible for the THz AFM magnons to carry a coherent GHz spin current. Here the instantaneous ac spin current would be transmitted as if it were a dc spin current, provided that the instantaneous spin orientation has a finite component parallel to the spin axis of the THz magnons. This condition is more likely to occur in a more isotropic AFM insulator (e.g., CoO and NiO) near the Néel temperature. However for AFM insulators with a strong uniaxial anisotropy, where the spin axes of all magnons lie along the same direction, we would expect a spin current with spin orientation perpendicular to the AFM spin axis to be filtered out [28]. This picture is consistent with our observation that the ac spin current transmission in CoO behaves in a similar way to the dc spin current transmission, with an enhancement around the Néel temperature. Our results suggest the need for more theoretical work to explore these mechanisms and to address this issue quantitatively.

In summary, we have measured spin precession using element- and time-resolved XMCD and XMLD in Py/Ag/CoO/Ag/Fe$_{75}$Co$_{25}$/MgO(001), using precessional pumping of the Py to generate a coherent GHz ac spin current. We find that the Fe$_{75}$Co$_{25}$ spins can be driven coherently through the AFM CoO by the GHz ac spin current with a peak in the precession amplitude around the CoO Néel temperature. In contrast, no GHz ac XMLD signal was observed from the CoO, suggesting that transmission of the spin current through the AFM CoO is not mediated by evanescent GHz frequency waves.

**Methods**
**X-ray pump-probe measurements**

X-ray pump-probe measurements were performed at Beamline 4.0.2 of the Advanced Light Source (ALS) at Lawrence Berkeley National Laboratory (LBNL). Microwave current of 4 GHz frequency is delivered to a CPW that generates a rf field at the sample. The rf excitation is synchronized to the ~500MHz electron bunch frequency of the storage ring, to ensure a fixed phase relation between the microwave pump signal and the probing x-ray pulses. This enables a stroboscopic measurement of the excited magnetic moments. We carried out phase delay scans by incrementally changing the time delay of the rf field with respect to the x-ray probe pulses, enabling us to map out the magnetization precession and to obtain detailed information about the precession amplitude and phase. The CPW contains a small hole (diameter~0.5 mm) in the signal line, which allows transmission of x-rays to the sample without affecting the CPW performance. A photodiode collects the luminescence yield from the sample to obtain the XMCD/XMLD signal. The x-ray incidence angle was 50° relative to the surface normal of the sample so that the in-plane component of the spin precession excited by the CPW could be obtained by element-resolved XMCD measurements as a function of the time delay of the microwave rf field [23,24]. For the XMCD measurements, the x-ray energy was tuned to the Ni $L_3$ edge (852.5 eV) and the Co $L_3$ edge (778.2 eV) to observe the dynamic Py and Fe$_{75}$Co$_{25}$ XMCD signals, respectively. For the linear dichroism measurements, the photon energy was tuned to 778.8 eV, to obtain the maximum CoO XMLD effect.

**Acknowledgements**

We thank Dr. Satoru Emori and Professor Yuri Suzuki for their great help in the FMR characterization. This work was supported by US Department of Energy, Office of Science, Office of Basic Energy Sciences, Materials Sciences and Engineering Division under Contract No. DE-AC02-05-CH11231 (van der Waals heterostructures program, KCWF16), National Science Foundation Grant No. DMR-1504568, Future Materials Discovery Program through the National Research Foundation of Korea (No. 2015M3D1A1070467), Science Research Center Program through the National Research

Foundation of Korea (No. 2015R1A5A1009962), ERATO-SQR project, JST, Japan, and Engineering and Physical Science Research Council (EPSRC) grants (EP/J018767/1 and EP/P02047X/1). This research used resources of the Advanced Light Source, which is a DOE Office of Science User Facility under contract no. DE-AC02-05CH11231.

**Author contributions**

Q. L., M. Y. and Z. Q. Q. designed and performed the experiments, analyzed the data and wrote the paper. C. K., P. S., A. T. N., and E. A. performed the XFMR measurements and contributed to the discussion. D. H. and E. S. carried out the FMR characterization and contributed to the discussion. R. J. H. helped build the XFMR setup and contributed to the discussion. T. Y. W., N. G., C. H. and J. L. were involved in the analysis and discussion of the data. [+]Q. L. and M. Y. contributed equally to this work.

**Competing interests**

The authors declare no competing interests.

**Additional information**

[*]Correspondence and request for materials should be addressed to Z. Q. Qiu:qiu@berkeley.edu.